\documentclass[a4paper,11pt,dvitex]{article}
\usepackage{pos}

\def\simge{\mathrel{%
       \rlap{\raise 0.511ex \hbox{$>$}}{\lower 0.511ex \hbox{$\sim$}}}}
\def\simle{\mathrel{%
       \rlap{\raise 0.511ex \hbox{$<$}}{\lower 0.511ex \hbox{$\sim$}}}}

\title{First-order phase transitions in the heavy quark region of lattice QCD at high temperatures and high densities}
\ShortTitle{First-order phase transitions in the heavy quark region of lattice QCD at high $T$ and $\mu$}

\author*[a]{Shinji Ejiri}

\affiliation[a]{Department of Physics, Niigata University, Niigata 950-2181, Japan}

\emailAdd{ejiri@muse.sc.niigata-u.ac.jp}

\abstract{
If there is a first-order phase transition in the light quark region of 2+1-flavor finite temperature and density QCD and if the region of the first-order phase transition expands with increasing density as suggested by several studies, then, at very high densities, we may expect that the first-order phase transition region expands into the heavy quark region of QCD, where we can perform efficient large scale simulations by adopting an effective theory of heavy quark QCD based on the hopping parameter expansion.
In the heavy quark region of QCD, we have another first-order phase transition region around the heavy quark limit at zero density. By numerical simulations of effective heavy quark QCD, we found that, the first-order transition at zero density turns into a crossover as the chemical potential is increased, but, when we increase the chemical potential further, the change in the plaquette value near the crossover point becomes much steeper. This may be suggesting reappearance of the first-order phase transition.
In this talk, we first show the nature of the phase transition of phase-quenched finite density QCD in the heavy quark region and then study the effect of the complex phase to discuss whether the QCD phase transition changes again to a first-order phase transition at very high densities.
}

\FullConference{The 41st International Symposium on Lattice Field Theory (LATTICE2024)\\
 28 July - 3 August 2024\\
Liverpool, UK\\}

\begin{document}
\maketitle

\section{Introduction}
\label{sec:intro}

Searching for the critical point where the finite temperature phase transition of QCD changes from a crossover to a first-order phase transition at finite density is one of the most interesting topics in the study of QCD.
If we consider the mass of the dynamical quark to be a controllable parameter, there are first-order phase transition regions in the light and heavy quarks regions, and in the intermediate region there is a crossover transition at zero density.
We expect the first-order phase transition in the light mass region to expand with increasing density.
In this case, the density at which the physical mass point enters the first-order phase transition region from the crossover region is the QCD critical point at that finite density \cite{deForcrand:2006pv,Karsch:2003va,Ejiri:2012rr}.
And if the first-order transition region continues to expand with density, the first-order phase transition region may expand into the heavy quark region.
The sign problem in finite density QCD is serious near the physical mass point, but is less serious in the heavy quark region.
We believe that useful information regarding the boundaries of first-order phase transitions can be obtained in the heavy quark region, and discuss the appearance of first-order phase transitions in the heavy quark region.

In Refs.~\cite{Saito:2011fs,Saito:2013vja,Ejiri:2019csa,Kiyohara:2021smr,Ashikawa:2024njc}, we determined the critical points in the heavy quark region using an effective theory based on the hopping parameter expansion.
Expanding the quark determinant in terms of the hopping parameter $\kappa$, we found that the expansion terms represented by Wilson loops winding around the periodic boundary in the time direction are important, and 
the higher order expansion terms are very strongly correlated with the leading term: Polyakov loop \cite{Wakabayashi:2021eye,Ejiri:2023tdp}. 
Using these properties, we proposed an effective theory of QCD in the heavy quark region that includes a Polyakov loop term.
Simulations of this effective theory require only the same computational cost as quenched QCD, allowing for calculations with high statistics and large space volumes \cite{Kiyohara:2021smr,Ashikawa:2024njc}.
Furthermore, in the heavy quark region, it is found that the phase of the quark determinant at finite density can be obtained with good approximation from the imaginary part of the Polyakov loop~\cite{Ejiri:2023tdp}, making it possible to study finite density QCD with high statistics.

In this study, we explore the phase structure of finite density lattice QCD by increasing the chemical potential while keeping the quark masses large.
In the next section, we explain the effective theory in the heavy quark region. Then, in Sec.~\ref{sec:phquench}, we discuss the case where the effect of the complex phase in finite density QCD is ignored.
The effect of complex phase is discussed in Sec.~\ref{sec:density}.
Section \ref{sec:summary} presents conclusions and future prospects.

\section{Effective theory in the heavy quark region}
\label{sec:effective}

We stidy $N_{\rm f}$-flavor QCD with the standard plaquette gauge action and the Wilson quark action on 
$N_s^3 \times N_t$ lattices. $N_{\rm site}=N_s^3 \times N_t$.
The partition function is given by 
\begin{equation}
Z= \int {\cal D} U \, \prod_{f=1}^{N_{\rm f}} 
\det M(\kappa_f, \mu) \, e^{6 N_{\rm site} \beta P} , 
\label{eq:z}
\end{equation}
where $P$ the plaquette value, and
$M$ is the Wilson quark kernel, which is a function of the chemical potential $\mu$ and the hopping parameter $\kappa$.
The hopping parameter is approximately proportional to the inverse quark mass.
In the heavy quark region, we evaluate the quark determinant by the hopping parameter expansion around $\kappa=0$:
\begin{eqnarray}
\ln \det M(\kappa) = 
\ln \det M(0) + N_{\rm site} \sum_{n=1}^{\infty} D_{n} \kappa^{n} , 
\hspace{5mm}
D_n 
=  \frac{(-1)^{n+1}}{N_{\rm site} \ n} \; {\rm tr} 
\left[ \left( \frac{\partial M}{\partial \kappa} \right)^n \right] .
\label{eq:tayexp}
\end{eqnarray}
The first term is $\ln \det M(0) =0$, and the matrix elements of 
$(\partial M/\partial \kappa)_{xy}$ are the hopping terms in $M$.
Nonzero contributions appear when the product of the hopping terms form closed loops in the space-time. 
Therefore, the nonzero contributions to $D_n$ are given by $n$-step Wilson loops and Polyakov loops. 
The latter are closed by the boundary condition, where we impose the anti-periodic boundary condition in the time direction for fermions.
These $D_n$ are classified by the number of windings $m$ in the time direction:
\begin{eqnarray}
D_n = W(n) + \sum_{m=1}^{\infty} L_m^+ (N_t, n) e^{m \mu/T} + \sum_{m=1}^{\infty} L_m^- (N_t, n) e^{-m \mu/T}
\label{eq:loopex}
\end{eqnarray}
$W(n)$ is the sum of Wilson-loop-type terms with winding number zero. $L_m^+ (N_t, n)$ and   $L_m^- (N_t, n)$ are the sum of Polyakov-loop-type terms with winding number $m$ in the positive and negative directions, respectively, and $L_m^- (N_t, n)=[L_m^+ (N_t, n)]^*$.
The lowest order term is $L_m^+ (N_t, mN_t)$.
In particular, $L_1^+ (N_t, N_t)$ is the usual Polyakov loop operator $\Omega$.

These expansion coefficients have been calculated on configurations generated near the phase transition point in Ref.~\cite{Wakabayashi:2021eye} and found to have the following properties: 
(1)~$W(n)$ mainly effect to shift the gauge coupling $\beta$, and have almost no effects in the determination of the critical $\kappa$.
(2)~$L_m (N_t, n)$ for $m \geq 2$ are much smaller than $L_1 (N_t, n)$. 
(3)~$L(N_t, n)$  is strongly correlated with the Polyakov loop ${\rm Re} \Omega$ on each configuration:
\begin{eqnarray}
L(N_t, n) \approx L^0 (N_t, n) c_n {\rm Re} \Omega,
\label{eq:lnapp}
\end{eqnarray}
where $L(N_t, n) = \sum_m [L_m^+ (N_t, n)+ L_m^- (N_t, n)] =  2\sum_m {\rm Re} L_m^+ (N_t, n)$, $c_n$ is a proportionality constant, and $L^0(N_t, n)$ is $L(N_t, n)$ when all link fields are set to $U_{x, \mu}=1$, which is given in Table~2 of Ref.~\cite{Wakabayashi:2021eye}.
(4)~The complex phases of $L_1^+ (N_t, n)$ and $\Omega$ are approximately equal \cite{Ejiri:2023tdp}:
\begin{eqnarray}
{\rm Arg} L_1^+ (N_t, n) \approx {\rm Arg} \Omega.
\label{eq:argapp}
\end{eqnarray}
Using these properties, we propose an effective action:
\begin{eqnarray}
S_{\rm eff} = -6N_{\rm site} \beta^* P - \frac{N_s^3 \lambda}{2} 
\left( e^{\mu/T} \Omega + e^{-\mu/T} \Omega^* \right), 
\hspace{5mm} 
\lambda = N_t \sum_{f=1}^{N_{\rm f}} \sum_{n=N_t}^{n_{\rm max}} L^0(N_t, n) \kappa_f^{n} c_n,
\label{eq:seff}
\end{eqnarray}
with which the partition function is given by $Z= \int {\cal D} U \, e^{-S_{\rm eff}}$.
$W(n)$ effectively shift $\beta$ from the original $\beta$ to 
$\beta^* = \beta + (1/6) W^0(4) \sum_f \kappa_f^4+ \cdots$ \cite{Wakabayashi:2021eye}.
Even if the number of expansion terms increases significantly, the effects of higher-order terms can be incorporated.
The computational cost of this simulation using $S_{\rm eff}$ is reduced to the same level as quenched QCD.

\section{Phase-quenched finite density QCD. }
\label{sec:phquench}

If we ignore the complex phase of $e^{-S_{\rm eff}}$, the effective action becomes
$S_{\rm eff} = -6N_{\rm site} \beta^* P - N_s^3 \lambda \cosh (\mu/T) \Omega_{\rm R}$, 
where $\Omega_{\rm R}$ is the real part of $\Omega$. 
Thus, by simply replacing $\lambda$ with $\lambda \cosh (\mu/T)$ from results calculated at $\mu =0$, we can investigate the phase-quenched finite density QCD, ignoring the phase, i.e., QCD with isospin chemical potential for $N_{\rm f}=2$.
Increasing $\lambda$ means increasing $\kappa$, so increasing $\lambda$ makes the approximation of the hopping parameter expansion worse.
Therefore, we fix $\kappa$ at a small value and increase $\mu/T$.
If we fix $\kappa$, the convergence does not worsen. 
For example, at $\lambda=0.005$ for $N_{\rm f}=2, N_t=6$, the convergence of the hopping parameter expansion is good. 
This point is in the crossover region, since the critical point at $\mu=0$ is $\lambda_c \approx 0.0013$.
Hereafter, changing the parameter $\lambda \cosh (\mu/T)$ will be considered to mean changing $\mu$ while ignoring the complex phase.

\begin{figure}[tb]
\centering
\vspace{-2mm}
\includegraphics[width=7.0cm,clip]{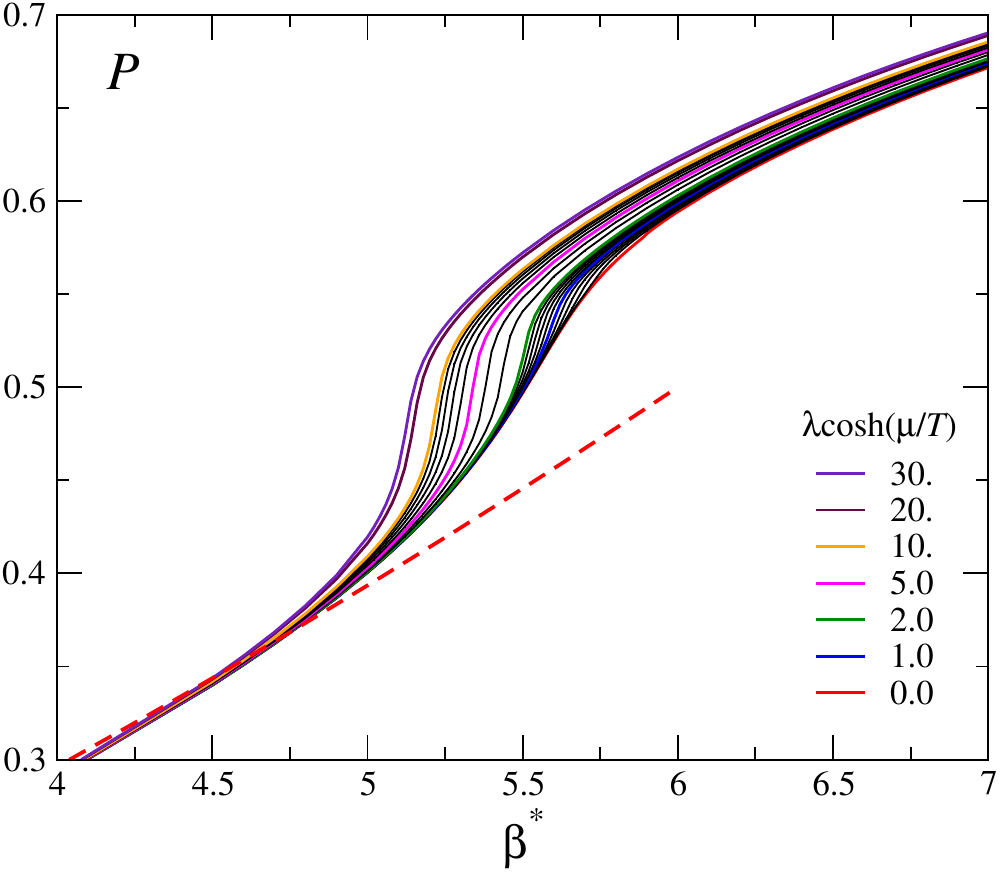}
\hspace{2mm}
\includegraphics[width=7.0cm,clip]{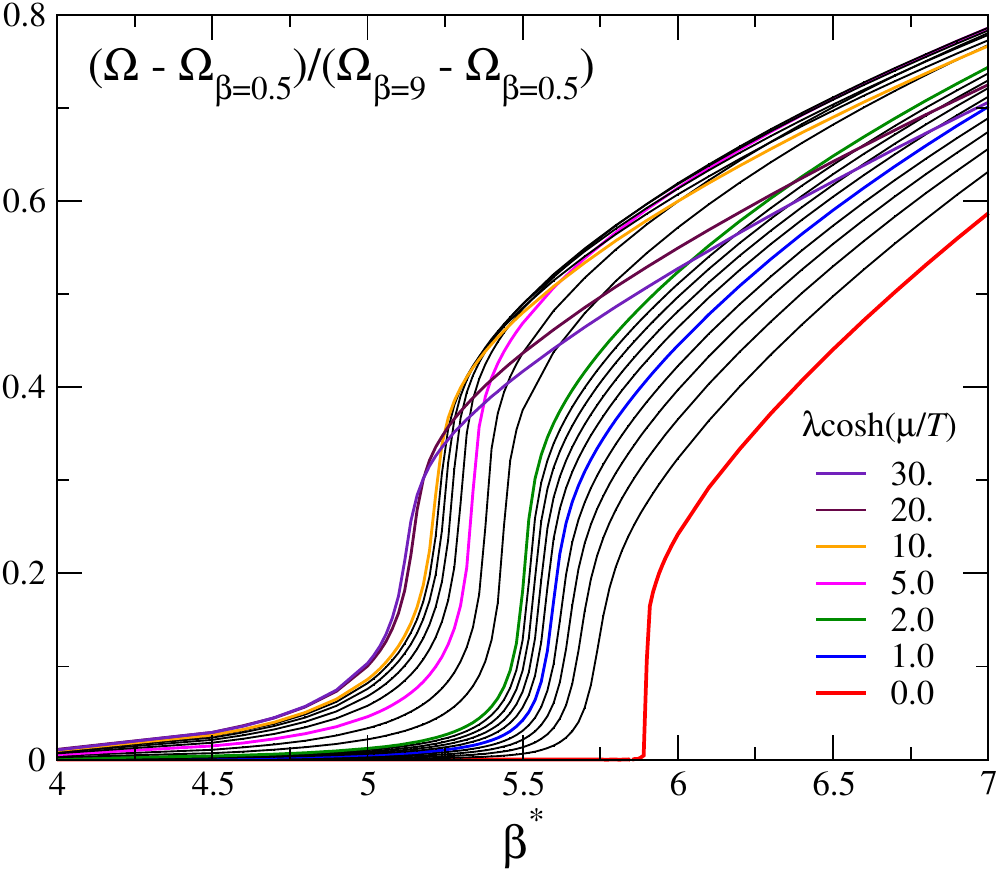}
\vspace{-1mm}
\caption{Left: Plaquette $\langle P \rangle$ as a function of $\beta^*$ for each $\lambda \cosh(\mu/T)$.
The dashed line is the strong coupling value.
Right: $(\langle \Omega_{\rm R} \rangle-\langle \Omega_{\rm R} \rangle_{\beta^*=0.5})/(\langle \Omega_{\rm R} \rangle_{\beta^*=9}-\langle \Omega_{\rm R} \rangle_{\beta^*=0.5})$ as a function of $\beta^*$ for each $\lambda \cosh(\mu/T)$.
}
\label{fig1}
\end{figure}

By performing simulations over a wide range of $\lambda \cosh(\mu/T)$, we investigate the phase-quenched finite density QCD.
The lattice size of the simulations $N_{\rm site}$ is $30^3 \times 6$.
The number of configurations for measurements is 10000 for each $\beta^*$ and $\lambda$.
The pseudo-heat-bath algorithm is adopted. 
The left panel of Fig.~\ref{fig1} is the expectation value of the plaquette $\langle P \rangle$ as a function of $\beta^*$ for each $\lambda \cosh(\mu/T)$.
When $\lambda \cosh(\mu/T) = 0$, the phase transition is first order, but imposing $\mu$ causes the change in $\langle P \rangle$ to become milder.
And from this figure, we can see that as $\mu$ is further increased, the change in $\langle P \rangle$ becomes steeper.
This result is reasonable.
The $\lambda \cosh(\mu/T)$ term in $S_{\rm eff}$ does not affect the calculation of $\langle P \rangle$ using the strong coupling expansion.
The read dashed line is the strng coupling value: $\langle P \rangle = \beta/18 +\beta^2/216$.
In the confinement phase, $\langle P \rangle$ is consistent with the result by the strong coupling
expansion.
Since the $\lambda \cosh(\mu/T)$ term forces the phase to become the deconfinement phase even in the region where $\beta$ is small, the change in $\langle P \rangle$ increases as $\mu$ increases at the phase boundary.
Furthermore, in two-flavor QCD, ignoring the complex phase is equivalent to imposing an isospin chemical potential, so increasing $\mu$ is considered to approach a pion condensed phase \cite{Brandt:2017oyy}.
In this respect as well, it is natural that $\langle P \rangle$ changes more sharply at the crossover point with increasing $\mu$.

The expectation value of the Polyakov loop $\langle \Omega_{\rm R} \rangle$ also changes more rapidly as $\mu$ increases.
We plot 
$(\langle \Omega_{\rm R} \rangle-\langle \Omega_{\rm R} \rangle_{\beta^*=0.5})/(\langle \Omega_{\rm R} \rangle_{\beta^*=9}-\langle \Omega_{\rm R} \rangle_{\beta^*=0.5})$
as a function of $\beta^*$ for each $\lambda \cosh(\mu/T)$ in the right panel of Fig.~\ref{fig1}.
Since $\langle \Omega_{\rm R} \rangle$ becomes a large value as $\lambda$ increases even in the region where $\beta$ is small, we subtracted the value of $\langle \Omega_{\rm R} \rangle$ at $\beta=0.5$ to focus on the change in $\langle \Omega_{\rm R} \rangle$ at the transition point.
Except for $\lambda=0$, as $\lambda \cosh(\mu/T)$ increases, the Polyakov loop also changes more steeper.
$\langle \Omega_{\rm R} \rangle$ changes almost perpendicular to the horizontal axis for large $\mu$.
It is interesting to see whether these quantities change more rapidly in finite density QCD, which incorporates the effects of the complex phase, and whether they behave like a first-order phase transition.

\section{Complex phase effect  at finite density}
\label{sec:density}

\begin{figure}[tb]
\centering
\vspace{-2mm}
\includegraphics[width=7.4cm,clip]{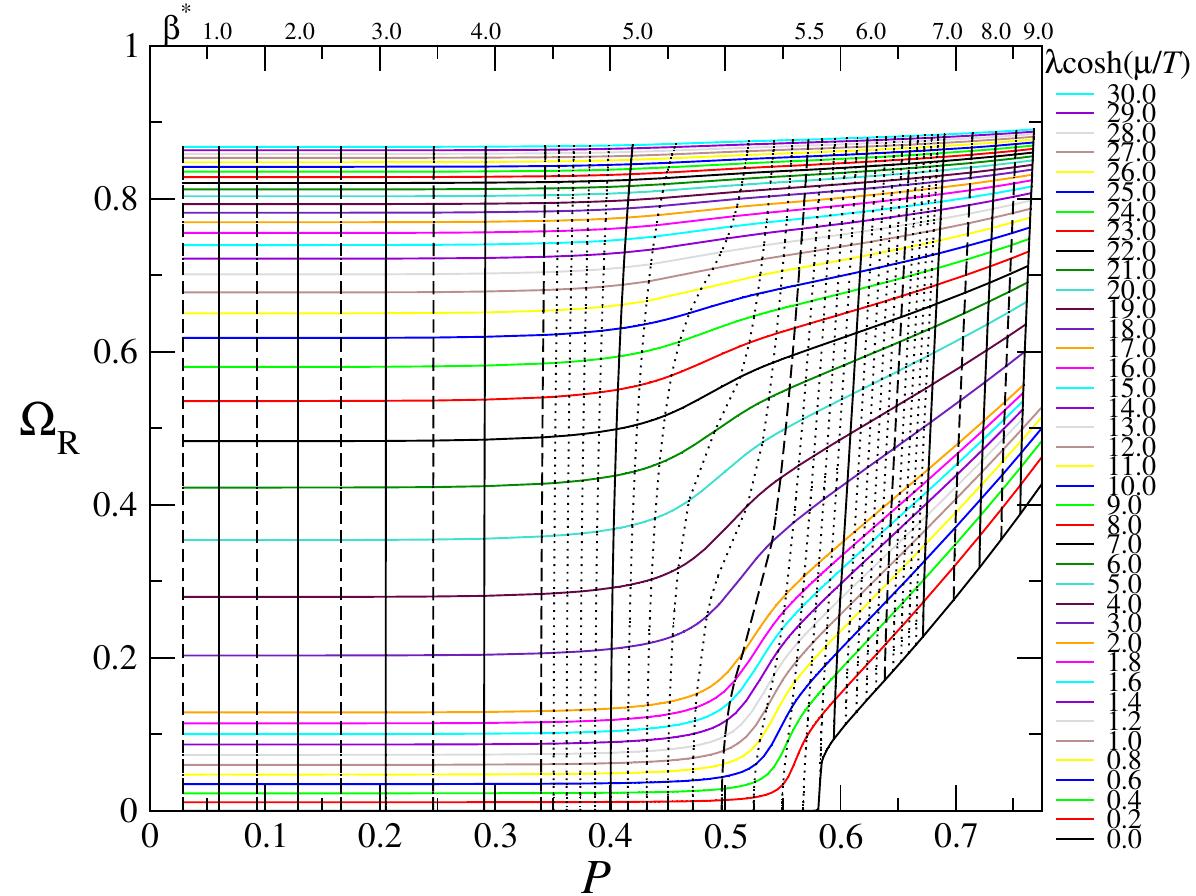}
\hspace{0mm}
\includegraphics[width=7.4cm,clip]{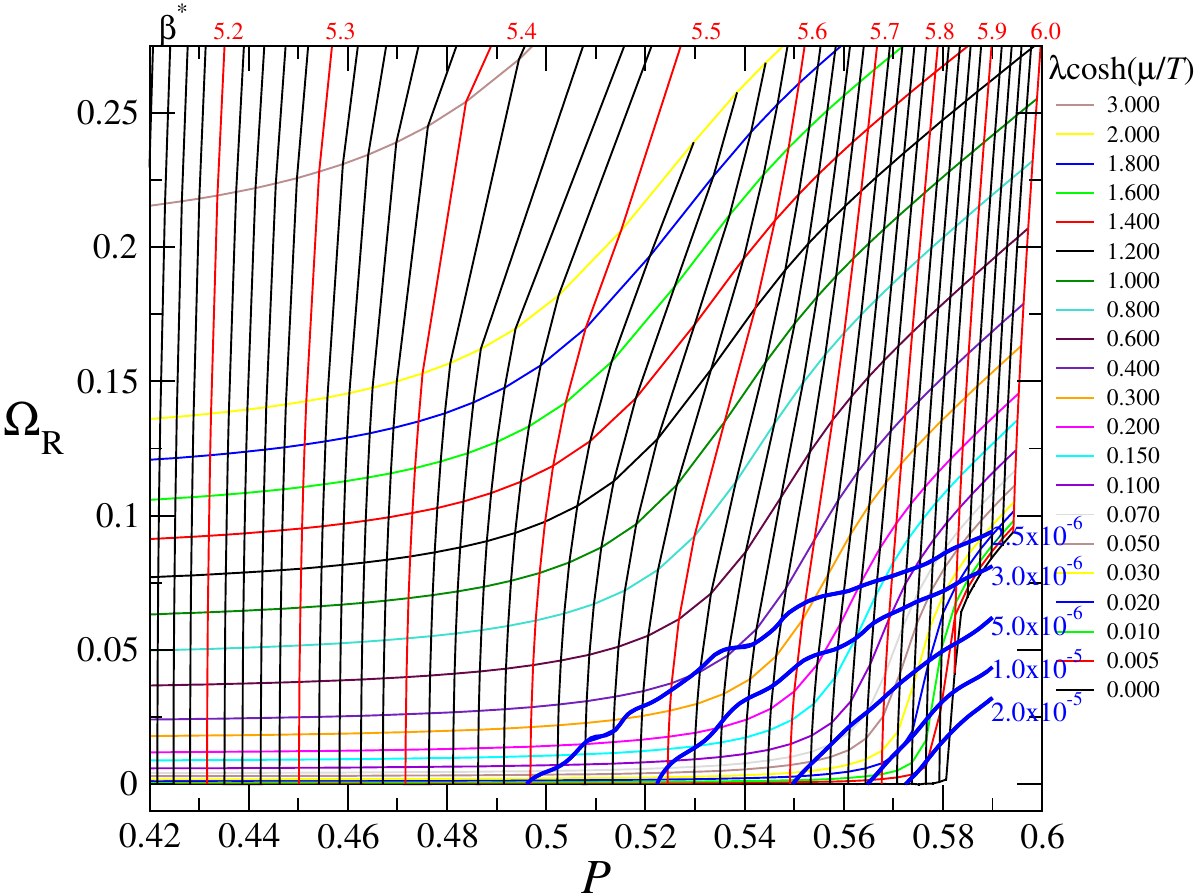}
\vspace{-1mm}
\caption{The left figure shows the parameters $(\beta, \lambda)$ whose expectation values are $(\langle P \rangle, \langle \Omega_{\rm R} \rangle)$when $\mu=0$.
The right figure is an enlarged view.
The bold blue lines are the contour lines of $\langle \Omega_{\rm I}^2\rangle$.
}
\label{fig2}
\end{figure}

Introducing the probability distribution function for $(P, \Omega_{\rm R})$, $W(P, \Omega_{\rm R})$, we rewrite the partition function as follows:
\begin{eqnarray}
Z &=& \int W(P, \Omega_{\rm R}) e^{6N_{\rm site} \beta^* P} 
e^{N_s^3 \lambda \cosh \frac{\mu}{T} \ \Omega_{\rm R}}
\left\langle \cos\left( N_s^3 \lambda \sinh \frac{\mu}{T} \ \Omega_{\rm I} \right) \right\rangle_{P, \Omega_{\rm R}} dP d\Omega_{\rm R}
\nonumber \\
& \equiv& \int F(P, \Omega_{\rm R}) dPd\Omega_{\rm R} , 
\label{eq:zfd}
\end{eqnarray}
where $\langle \cdots \rangle_{P, \Omega_{\rm R}}$ means the expectation value with fixed $(P, \Omega_{\rm R})$.
For the case of $\mu=0$, when configurations are generated at $\beta^*$ and $\lambda$, the following equations are satisfied at $(P, \Omega_{\rm R})$ where the configuration generation probability $F(P, \Omega_{\rm R})$ is maximized:
\begin{eqnarray}
\frac{\partial \ln F(P, \Omega_{\rm R})}{\partial P} = 6N_{\rm site} \beta^* + \frac{\partial \ln W(P, \Omega_{\rm R})}{\partial P} =0, \hspace{2mm}
\frac{\partial \ln F(P, \Omega_{\rm R})}{\partial \Omega_{\rm R}} = N_s^3 \lambda + \frac{\partial \ln W(P, \Omega_{\rm R})}{\partial \Omega_{\rm R}} =0. 
\label{eq:probmax}
\end{eqnarray}
The peak position of $F(P, \Omega_{\rm R})$ is approximately equal to the expectation value $(\langle P \rangle, \langle \Omega_{\rm R} \rangle)$.
Figure~\ref{fig2} (left) shows lines of constant simulation parameters $\beta^*$ and $\lambda$ plotted on the $(\langle P \rangle, \langle \Omega_{\rm R} \rangle)$ plane.
The black lines roughly parallel to the vertical axis are the constant $\beta^*$ lines, and the values of $\beta^*$ are shown on the top of the figure.
The constant $\lambda$ lines are shown as colored lines, and the values of $\lambda$ are written on the right side of the figure.
Using Eq.~(\ref{eq:probmax}) and the information of Fig.~\ref{fig2}, $\partial \ln W(P, \Omega_{\rm R})/\partial P$ and $\partial \ln W(P, \Omega_{\rm R})/\partial \Omega_{\rm R}$ can be obtained.
The right figure is an enlarged view of the area near the first-order phase transition.

\begin{figure}[tb]
\centering
\vspace{-2mm}
\includegraphics[width=7.0cm,clip]{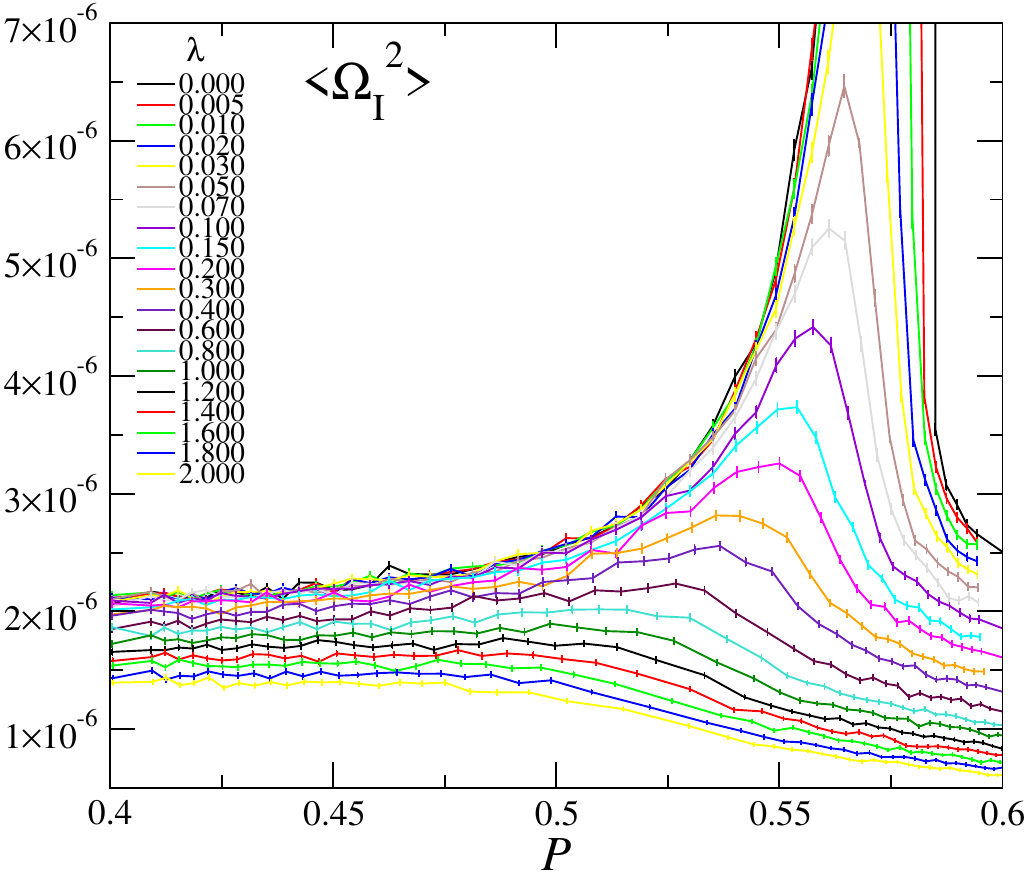}
\hspace{2mm}
\includegraphics[width=7.0cm,clip]{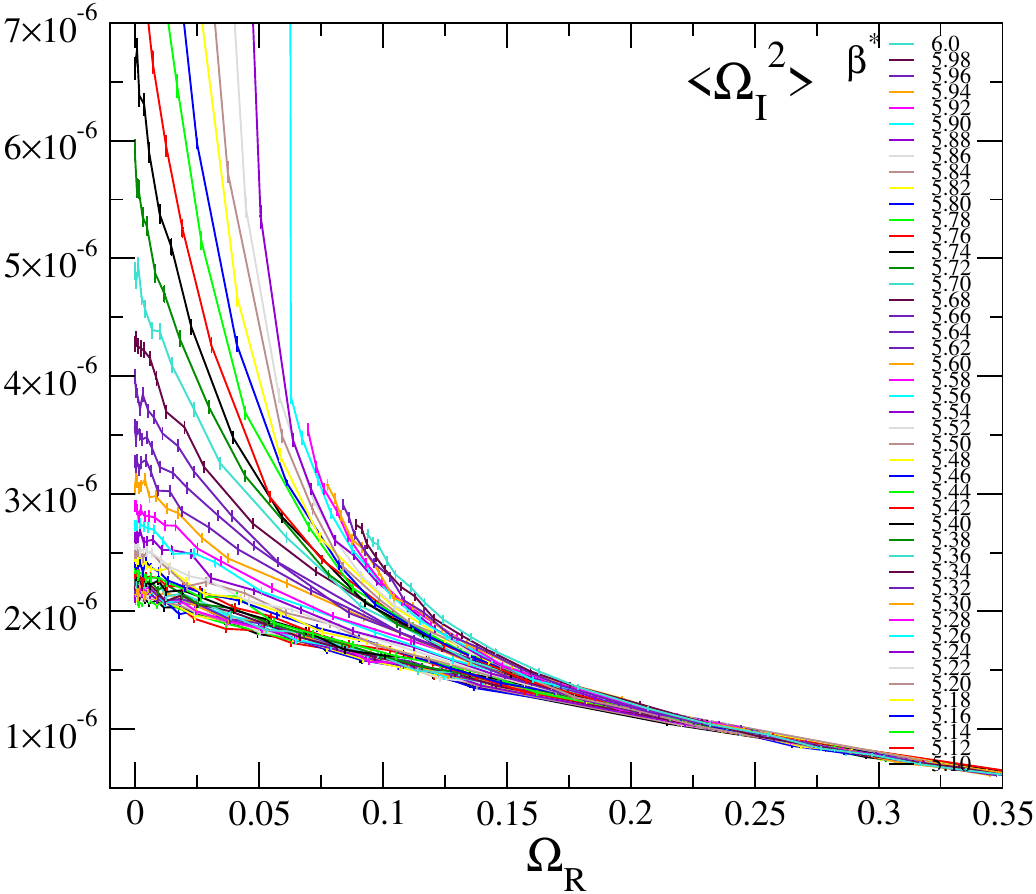}
\vspace{-1mm}
\caption{Left: $\langle \Omega_{\rm I}^2\rangle$ as a function of $P$ for each $\lambda$.
Right: $\langle \Omega_{\rm I}^2\rangle$ as a function of $\Omega_{\rm R}$ for each $\beta$.
}
\label{fig3}
\end{figure}

For finite $\mu$, the peak position of $F(P, \Omega_{\rm R})$ is given by
\begin{eqnarray}
\frac{\partial \ln F(P, \Omega_{\rm R})}{\partial P} & \! = & \!
6N_{\rm site} \beta^* + 
\frac{\partial \ln W(P, \Omega_{\rm R})}{\partial P} + 
\frac{\partial \ln \left\langle \cos\left( N_s^3 \lambda \sinh \frac{\mu}{T} \ \Omega_{\rm I} \right) \right\rangle_{P, \Omega_{\rm R}}}{\partial P}
=0, \nonumber \\
\frac{\partial \ln F(P, \Omega_{\rm R})}{\partial \partial \Omega_{\rm R}} & \! = & \! 
N_s^3 \lambda \cosh \frac{\mu}{T} + 
\frac{\partial \ln W(P, \Omega_{\rm R})}{\partial \Omega_{\rm R}} + 
\frac{\partial \ln \left\langle \cos\left( N_s^3 \lambda \sinh \frac{\mu}{T} \ \Omega_{\rm I} \right) \right\rangle_{P, \Omega_{\rm R}}}{\partial \Omega_{\rm R}}
=0, 
\label{eq:probmaxfd}
\end{eqnarray}
where $\Omega_{\rm I}$ is the imaginary part of $\Omega$. 
However, when $\mu$ is increased, a sign problem occurs in which 
$\langle \cos\left( N_s^3 \lambda \sinh \frac{\mu}{T} \ \Omega_{\rm I} \right) \rangle$ 
becomes zero.
To avoid the sign problem, we perform a cumulant expansion: 
$\ln \left\langle \cos\left( N_s^3 \lambda \sinh \frac{\mu}{T} \ \Omega_{\rm I} \right) \right\rangle
= -\frac{1}{2} \left\langle \left( N_s^3 \lambda \sinh \frac{\mu}{T} \ \Omega_{\rm I} \right)^2 \right\rangle_c
+ \frac{1}{4!} \left\langle \left( N_s^3 \lambda \sinh \frac{\mu}{T} \ \Omega_{\rm I} \right)^4 \right\rangle_c
- \frac{1}{6!} \left\langle \left( N_s^3 \lambda \sinh \frac{\mu}{T} \ \Omega_{\rm I} \right)^6 \right\rangle_c + \cdots$. 
Here, 
$\langle x^2 \rangle_c = \langle x^2 \rangle, 
\langle x^4 \rangle_c = \langle x^4 \rangle -3 \langle x^2 \rangle^2, 
\langle x^6 \rangle_c = \langle x^6 \rangle -15 \langle x^4 \rangle \langle x^2 \rangle^2 +30 \rangle \langle x^2 \rangle^3, 
\cdots $.
We approximate 
\begin{eqnarray}
\ln \left\langle \cos\left( N_s^3 \lambda \sinh \frac{\mu}{T} \Omega_{\rm I} \right) \right\rangle_{P, \Omega_{\rm R}} \approx -\frac{1}{2} \left\langle \left( N_s^3 \lambda \sinh \frac{\mu}{T} \Omega_{\rm I} \right)^2 \right\rangle.
\end{eqnarray}
We believe this approximation is sufficient for a qualitative estimation.
We call it the Gaussian approximation because the equation holds strictly when the complex phase is Gaussian distributed~\cite{Ejiri:2007ga,Ejiri:2009hq}.
Since the distributions of $P$ and $\Omega_{\rm R}$ are narrow in each simulation, the difference between $\langle \Omega_{\rm I}^2\rangle_{P, \Omega_{\rm R}}$ with a fixed $P$ and $\Omega_{\rm R}$ and the usual expectation value $\langle \Omega_{\rm I}^2\rangle$ is within the error.
Using Eq.~(\ref{eq:probmax}), the derivatives of $\ln W(P, \Omega_{\rm R})$ can be measured by $\mu=0$ simulations. We denote the simulation parameters at $\mu=0$ as $(\beta^*_0, \lambda_0)$.
The parameters $\beta^*$ and $\lambda$ for which the point where $F(P, \Omega_{\rm R})$ is maximum at finite $\mu$ is $(P, \Omega_{\rm R})$ can be calculated using the following equations:
\begin{eqnarray}
\beta^* = \beta^*_0 + \frac{N_s^3}{12N_t} \left( \lambda \sinh \frac{\mu}{T} \right)^2 
\frac{\partial \langle \Omega_{\rm I}^2 \rangle}{\partial P}, \hspace{5mm} 
\lambda \cos \frac{\mu}{T} = \lambda_0 + \frac{N_s^3}{2} \left( \lambda \sinh \frac{\mu}{T} \right)^2 
\frac{\partial \langle \Omega_{\rm I}^2 \rangle}{\partial \Omega_{\rm R}}.
\label{eq:fdbetalambda}
\end{eqnarray}
If we equate the expectation value of $P$ and $\Omega_{\rm R}$ with the point with the maximum probability of generation,
these equations give us $(\beta^*, \lambda)$ at finite $\mu$ whose expectation values, $\langle P \rangle$ and $\langle \Omega_{\rm R} \rangle$, are the same as the expectation values of zero density simulations.
The calculation of $\langle \Omega_{\rm I}^2\rangle$ is relatively easy.
We plot it in Fig.~\ref{fig3}.
The left panel shows $\langle \Omega_{\rm I}^2\rangle$ as a function of $P$ for each $\lambda$, and the right panel shows $\langle \Omega_{\rm I}^2\rangle$ as a function of $\Omega_{\rm R}$ for each $\beta$.
However, the value of $\langle \Omega_{\rm I}^2 \rangle$ is large only near the first-order transition point at $\lambda=0$.
The blue bold lines in Fig.~\ref{fig3} (right) are the contour lines for 
$\langle \Omega_{\rm I}^2 \rangle = 2.5 \times 10^{-6}, 3.0 \times 10^{-6}, 5.0 \times 10^{-6}, 1.0 \times 10^{-5}$, and $2.0 \times 10^{-5} $.
This large peak of $\langle \Omega_{\rm I}^2 \rangle$ does not contribute to changing the nature of the phase transition at large $\mu$.

\begin{figure}[tb]
\centering
\vspace{-2mm}
\includegraphics[width=7.1cm,clip]{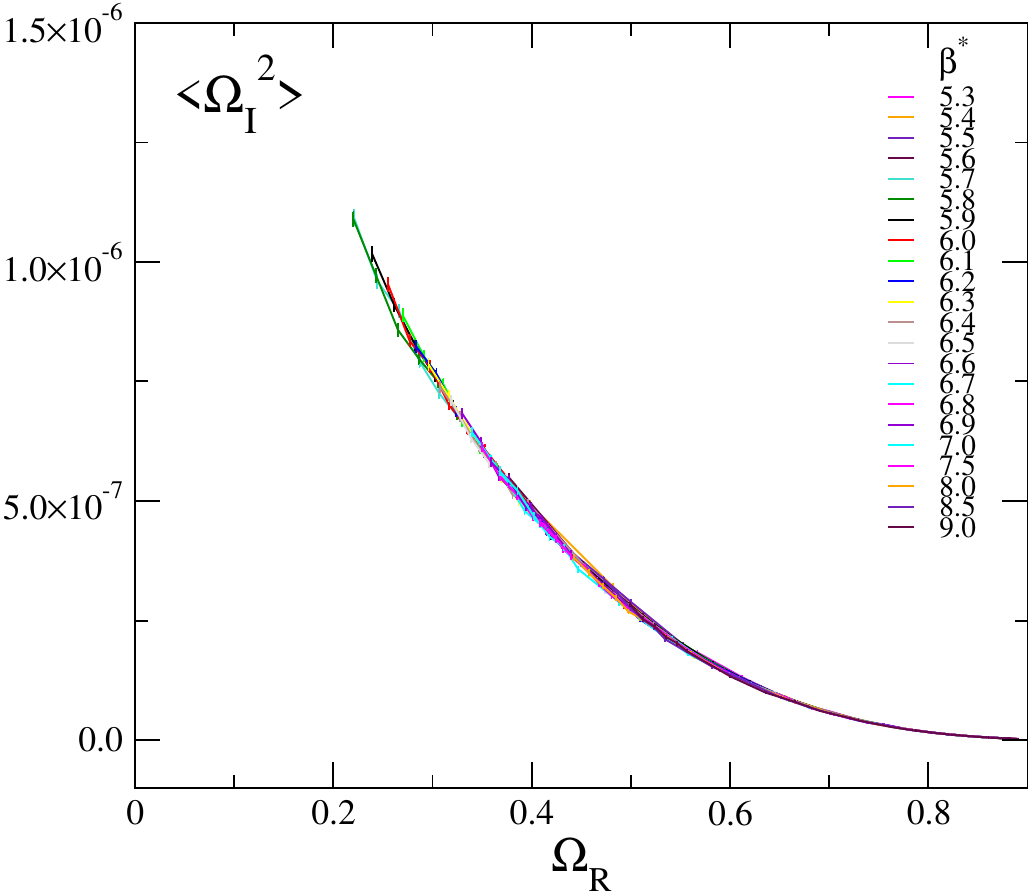}
\hspace{2mm}
\includegraphics[width=7.2cm,clip]{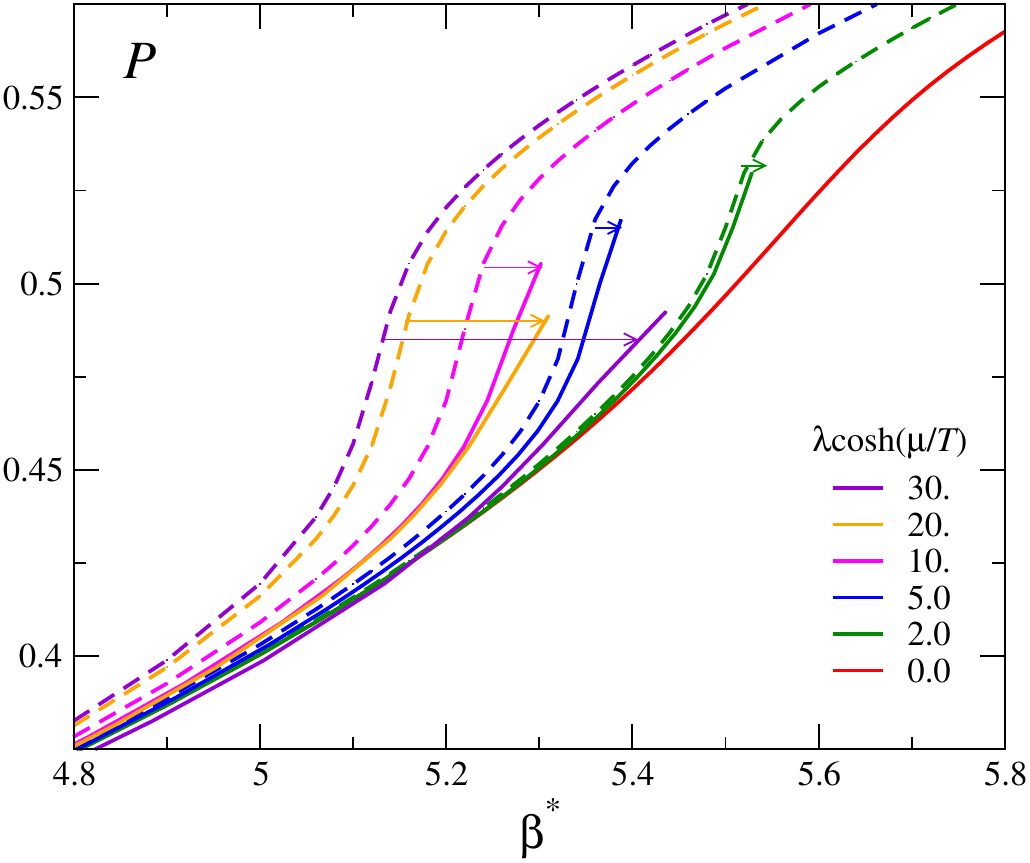}
\vspace{-1mm}
\caption{Left: Only the data of $\langle \Omega_{\rm I}^2\rangle$ in the high-temperature phase are plotted. 
Right: Plaquette as a function of $\beta$ without the complex phase effects (dashed line) and with complex phase effects (solid line).
}
\label{fig4}
\end{figure}

We estimate the change in $\beta^*$ by the finite density effect, which is given by 
$\partial \langle \Omega_{\rm I}^2 \rangle / \partial P$ in Eq.~(\ref{eq:fdbetalambda}).
The $P$ dependence of $\langle \Omega_{\rm I}^2\rangle$ is much smaller than $\Omega_{\rm R}$ dependence.
Especially in the high-temperature phase, no $P$ dependence is seen.
In the left panel of Fig.~\ref{fig4}, we plot only the data of $\langle \Omega_{\rm I}^2\rangle$ in the high-temperature phase and $\lambda \geq 1$. 
These are depend only on $\Omega_{\rm R}$.
This mean that
$\partial \langle \Omega_{\rm I}^2 \rangle/\partial P |_{\Omega_{\rm R}} =0$
in the high-temperature phase.
In the low-temperature phase, $P$ dependence is observed, but compared to its error, it is very small at large $\lambda$, making numerical calculation of 
$\partial \langle \Omega_{\rm I}^2 \rangle/\partial P |_{\Omega_{\rm R}}$
difficult.
On the other hand, the strong coupling analysis gives us its approximate value for small $\beta^*$.

We calsculate the strong coupling limit of $\langle \Omega_{\rm I}^2 \rangle$ at $\lambda=0$ for $N_c =3$.
\begin{eqnarray}
\langle \Omega_{\rm I}^2 \rangle \! &=& \! \left\langle \left(\frac{\Omega - \Omega^*}{2i} \right)^2 \right\rangle
\approx \frac{1}{2} \langle \Omega \Omega^* \rangle \nonumber \\
&& \hspace{-13mm} \approx \ \frac{N_s^3}{2N_s^6 N_c^2}\left( \left\langle {\rm tr}(UU \cdots U)_{\vec{x}} {\rm tr}(U^{\dagger} U^{\dagger} \cdots U^{\dagger})_{\vec{x}} \right\rangle
+6 \left\langle {\rm tr}(UU \cdots U)_{\vec{x}} {\rm tr}(U^{\dagger} U^{\dagger} \cdots U^{\dagger})_{\vec{x}+\hat{1}} \right\rangle + \cdots \right) \nonumber \\
&=& \frac{1}{2N_s^3 N_c^2} + \frac{3}{N_s^3 N_c} \left( \frac{\beta}{2 N_c^2} \right)^{N_t} .
\label{eq:pl2strong}
\end{eqnarray}
Here, the plaquette is given by $P= \beta/(2N_c^2)$ in the leading order.
Then, 
\begin{eqnarray}
\langle \Omega_{\rm I}^2 \rangle 
\approx \frac{1}{2N_s^3 N_c^2} + \frac{3}{N_s^3 N_c} P^{N_t}, 
\hspace{5mm}
\frac{\partial \langle \Omega_{\rm I}^2 \rangle}{\partial P}
\approx \frac{3 N_t}{N_s^3 N_c} P^{N_t-1}
\label{eq:dpl2strong}
\end{eqnarray}
This behavior is approximately consistent with the simulation data in the low-temperature phase, 
but $P$ dependence is very small.

To estimate how much $\beta^*$ shifts from $\beta^*_0$ for finite $\mu$, we assume 
\begin{eqnarray}
\left. \frac{\partial \langle \Omega_{\rm I}^2 \rangle}{\partial P} \right|_{\Omega_{\rm R}} =0 
\ \ {\rm at \ high} \ T \ {\rm and} \hspace{5mm} 
\left. \frac{\partial \langle \Omega_{\rm I}^2 \rangle}{\partial P} \right|_{\Omega_{\rm R}} =
\frac{3 N_t}{N_s^3 N_c} P^{N_t-1} (1- \Omega_{\rm R}) e^{-3 \Omega_{\rm R}}
\ \ {\rm at \ low} \ T.
\label{eq:betasft}
\end{eqnarray}
This function takes the value of the strong coupling limit of $\lambda=0$ at $\Omega_{\rm R}=0$ and zero at  $\Omega_{\rm R}=1$.
The ``3'' in $e^{-3 \Omega_{\rm R}}$ is chosen by eye to reproduce the simulation data.
The dashed lines in the right panel of Fig.~\ref{fig4} are the plaquette for each $\lambda \cosh(\mu/T)$ when the complex phase is ignored, which simply replaces $\lambda$ at $\mu=0$ with $\lambda \cosh(\mu/T)$.
The solid lines are the lines that incorporate the effect of the complex phase in the low-temperature phase by shifting the horizontal axis $\beta$ using Eq.~(\ref{eq:betasft}).
Since $\beta$ does not shift in the high-temperature phase, as $\mu$ increases, the $\beta$ shift increases the change in the plaquette at the phase transition point.
When $\mu$ is further increased, the difference between $\langle P \rangle$ of the low-temperature phase, in which $\beta$ shifts, and $\langle P \rangle$ of the high-temperature phase, in which $\beta$ does not shift, becomes larger, and a gap is expected to appear at the transition point.
This predicts the appearance of a first-order phase transition.
At the same time, $\lambda \cosh(\mu/T)$ also shifts, but we will not discuss this here, since $\lambda \cosh(\mu/T)$ changes the same way in both phases and is not related to the discontinuity.
For example, we adopt $\lambda =0.005$.
When $\lambda \cosh(\mu/T) =5.0$ or greater, the $\beta$ shift is significant, which corresponds to 
$\mu/T =7.60$.
Furthermore, when $\lambda \cosh(\mu/T) =10$, $\mu/T =8.29$ and $\lambda \cosh(\mu/T) =20$, $\mu/T =8.99$.
This suggests a discontinuity in $\langle P \rangle$ at the transition point, and a first-order phase transition is expected to appear at large $\mu/T$.

\section{Conclusions}
\label{sec:summary}

Even at finite density, the phase structure of QCD in the heavy quark region can be relatively easily studied by using an effective theory based on the hopping parameter expansion.
First, we discussed the nature of the phase transition of phase-quenched finite density QCD (QCD with an isospin chemical potential for $N_{\rm f}=2$) in the heavy quark region.
The first order transition at zero density turns into a crossover as $\mu$ is increased, but, when we increase $\mu$ further, the change in the plaquette value near the crossover point becomes much steeper.
Then, we estimate the effect of the complex phase of the quark determinant to discuss whether the QCD phase transition changes again to a first-order phase transition at very large $\mu$.
In the high temperature phase, the effect of the complex phase is negligible.
In the low temperature phase, the complex phase effect leads to a steeper change in the plaquette when estimated from the results in the strong coupling limit.
This effect of the complex phase becomes larger as $\mu$ increases.
This suggests the appearance of a first-order phase transition region at high density.

Since physical quantities change as a function of $\lambda \cosh(\mu/T)$, it is clear that the critical $\mu$ at which a first-order phase transition arises becomes smaller as $\kappa$ increases (the quark mass decreases).
The next question is whether the boundary line is connected to the boundary line of the first-order phase transition in the light quark region.
If connected, the critical $\mu$ at the heavy quarks provides important information for determining the critical $\mu$ at the physical quark masses.

\acknowledgments
The author would like to thank K. Kanaya and M. Kitazawa for discussions and comments, and for allowing use of the configuration generation code used in Refs.~\cite{Kiyohara:2021smr,Ashikawa:2024njc}.
This work was supported by JSPS KAKENHI Grant Numbers JP21K03550, JP19H05598.
This research used computational resources provided by the HPCI System Research project (Project ID: hp220020, hp220024), and SQUID at Cybermedia Center, Osaka University.

\end{document}